\documentclass[twocolumn,twoside,slac_two]{revtex4}
\usepackage{graphicx}
\usepackage{fancyhdr}
\begin{document}

\title{Turbulent Extreme Multi-Zone Model for Multi-Waveband Variations of Blazars}

%

\author{A.\ P.\ Marscher}
\affiliation{Institute for Astrophysical Research, Boston University,
725 Commonwealth Ave., Boston, MA 02215, USA}

\begin{abstract}
The author is developing a numerical code with thousands of emission zones
to simulate the time-dependent
multi-waveband emission from blazars. The code is based on a model in which
turbulent plasma flowing at a relativistic speed down a jet crosses a standing
conical collimation shock that accelerates electrons to maximum energies
in the 5-100 GeV range. This paper reports early results produced by the model.
The simulated light curves and time profiles of the
degree and position angle of polarization have a number
of features in common with the observational data of blazars. Maps of the
polarized intensity structure can be compared with those of
blazars observed with very long baseline interferometry at short millimeter wavelengths.
\end{abstract}

\maketitle

\thispagestyle{fancy}


\section{INTRODUCTION}

Reproduction of the time-dependent multi-waveband emission of blazars poses severe
theoretical challenges.
Observations with the Very Long Baseline Array (VLBA) at 7 mm wavelength \citep{mars12b}
have found that the majority of $\gamma$-ray flares in blazars are simultaneous with
either the passage of a new superluminal knot through the bright stationary feature termed
the ``core'' or a brightening of the core with no prominent knot subsequently appearing.
The core lies parsecs from the central engine in blazars \citep[see][for a discussion]{mars06}.
Nevertheless, time-scales of flux changes can be as short as hours, even for very high-energy
$\gamma$ rays that must also be emitted on parsec scales to avoid a high opacity to pair-production
\citep[e.g.,][]{alek11}.

One way to explain the rapid variability so far from the central engine is to imagine
that the emission arises from
localized regions where Doppler factors exceed the values derived from apparent superluminal
speeds of radio knots. Two possibilities to realize this scenario have been proposed:
magnetic reconnections \citep{gian09} and relativistic turbulence \citep{mj10,np12}. The author
is developing a numerical code, called the Turbulent Extreme Multi-Zone (TEMZ) model,
that can compute the time-dependent spectral energy distribution (SED)
and therefore multi-frequency light curve for these physical scenarios. Thus far, the code has been
applied to an approximate simulation of turbulent plasma crossing a standing conical shock in the jet.
It includes two sources of seed photons for inverse Compton scattering: a hot, dusty torus and
a Mach disk (also often called a ``working surface''), which is a strong shock oriented transverse
to the jet flow that occupies a small portion of the jet cross-section centered on the axis
\citep{cf48}. One of the primary merits of this effort is that the simulations can be compared with
the wealth of data produced by multi-waveband monitoring programs: light curves, SEDs, and
polarization vs. time, frequency, and position. This progress report presents early results from this effort.

\section{DESCRIPTION OF THE TEMZ MODEL}
\begin{figure*}[t]
\vspace{-1cm}
\includegraphics[width=150mm]{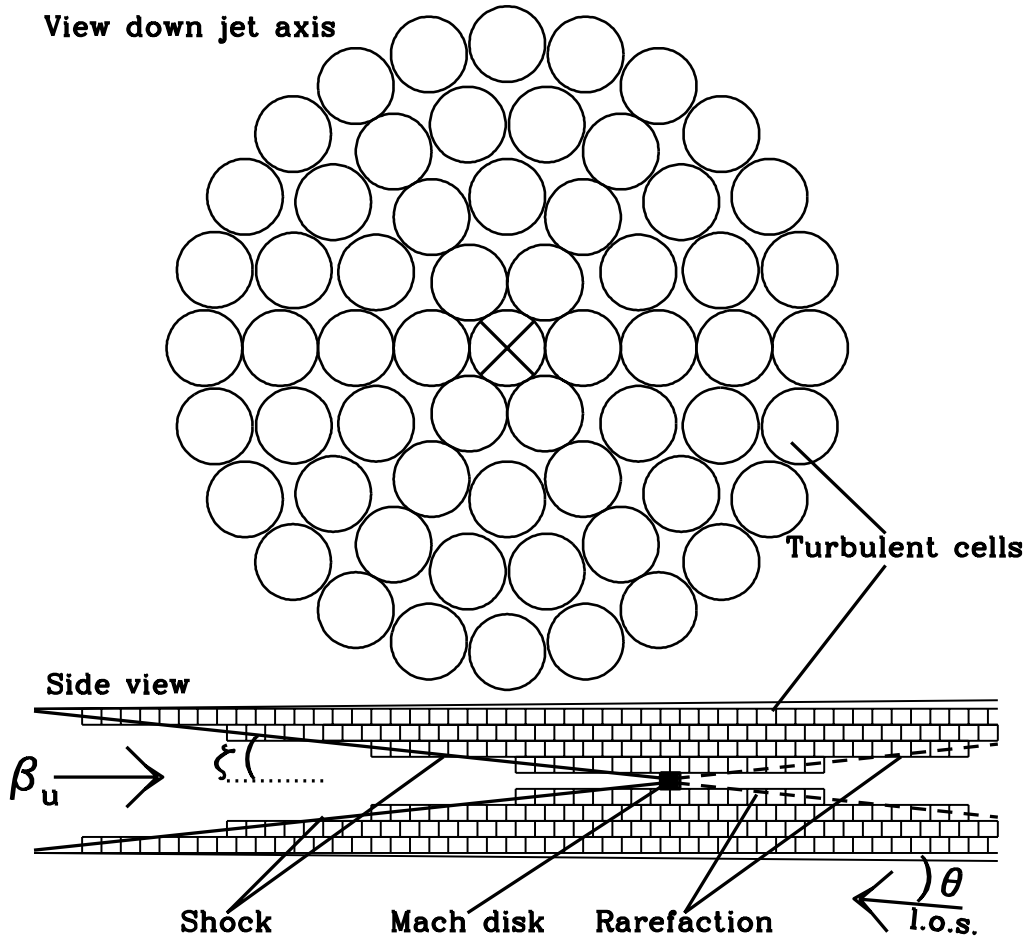}
\caption{Schematic of the TEMZ model showing the shock, rarefaction, and cylindrical cells.
The emission is assumed to occur between the conical standing shock and the rarefaction.
The only practical restriction on the number of cells $N_{\rm cross}$ across the jet
cross-section (as seen in view down axis on bottom) is the memory of the computer, although the
mean observed degree of linear polarization $\langle \Pi_{\rm opt} \rangle$ at optical wavelengths
provides a guideline: $N_{\rm cross} \sim 1.4\langle \Pi_{\rm opt} \rangle^{-1/2}$.
}
\label{fig1}
\end{figure*}

The code is based on a model in which turbulent plasma flowing at a relativistic speed down a jet
crosses a standing conical collimation shock that accelerates electrons to energies
up to some maximum value, probably in the range from 5 to 100 GeV.
The code divides the plasma into many cells (see Fig.\ \ref{fig1}),
each with a different direction of magnetic field. This provides a crude
simulation of turbulence. The energy density of electrons is
modulated randomly such that the variations in flux produce a power spectral
density described by a power law with a slope similar to the observed value
in the range of 1.2-2 \citep{Lars12}.
The direction of the magnetic field is different for each cell, and the
maximum electron energy generated at the shock can either depend on that direction
or be set randomly within a probability distribution. The plasma in each cell
has a turbulent velocity with random direction superposed on the systemic flow
velocity. As the plasma flows across the shock and then loses energy as it
advects downstream, variations in the synchrotron and inverse Compton flux,
as well as in the linear polarization, occur over a range of time-scales.
Since the standing shock lies parsecs from the central engine, radiation from a hot dust torus
and a Mach disk at the vertex of the shock cone can both be significant sources of
seed photons from the scattering. Emission from the latter varies along with the energy density of the
plasma, although the photons from an outburst require months to diffuse across the
cells; this smoothes the fluctuations of the seed photon field.
Traditional synchrotron self-Compton (SSC) scattering of the synchrotron
radiation emitted by each of the cells other than the Mach disk is not included in
the current, non-parallelized version of the code owing to the excessive cpu time needed.

The motivation for the model arises from salient observed properties of the multi-frequency
behavior of blazars. Most of these properties are ignored as ``weather'' in other models, but
are used by the author to both shape the TEMZ model and provide tests of its ability to represent
the physics of blazars. The critical observations include:
(1) red-noise power spectra of flux variations in blazars \cite[e.g.,][]{chat08,abdo10,chat12,Lars12},
(2) shorter time-scales of variability of flux and polarization at higher frequencies \citep{mars12a}, (3) mean polarization levels as well as fractional deviations from the mean that are higher at optical than at lower frequencies \citep{jor07,darc10}, (4) apparent rotations in polarization position angle that are really just random walks of the projected magnetic field direction \citep{darc07,darc10}, (5) breaks in the synchrotron spectrum by more than the radiative loss value of 0.5 \citep[e.g.,][]{wehrle12}, and (6) flares that are often sharply peaked or contain multiple peaks, neither of which is reproduced by single- or few-zoned models. The dependence of items 2-4 on frequency is directly related to the change in spectral index beyond the break, according to the model \citep{mj10}.

An important feature of the model is the different value of the maximum electron energy
$\gamma_{\rm 0,max}$ that is achieved
by particle acceleration at the shock front. This can either be determined randomly from within a
power-law probability distribution \citep{mj10} or as a function of the angle that the magnetic field
of the cell subtends to the shock normal. The simulations presented here adopt
$\gamma_{\rm 0,max}$ times $\cos^2(B_\parallel/B)$, where the subscript $\parallel$ indicates the component of the magnetic field that is parallel to the shock normal. The value of $\gamma_{\rm 0,max}$ is not
allowed to fall below some minimum value, which is set as an input parameter.
Future development can include results from Monte Carlo calculations
of particle acceleration in relativistic shocks \citep[e.g.,][]{sb12}.

The code currently includes synchrotron radiation, inverse Compton scattering (IC) of seed photons from a hot dust torus, and IC of synchrotron + SSC radiation from relatively slowly moving plasma in a Mach disk at the vertex of the conical shock. The last of these is abbreviated as ``SSC-MD.'' If a Mach disk is present, this can be the dominant SSC emission, since the seed photons are Doppler boosted in the frame of the turbulent cells. The combined effects of co-spatiality of the emission regions at different wavebands, non-uniform electron energy distribution, different magnetic field orientations for different turbulent cells, and light-travel delays, cause correlations of variations at pairs of wavebands but with time lags and often a lack of one-to-one correspondence of flares. This is similar to the observed behavior of blazars \cite[e.g.,][]{mars12a,wehrle12}. The code generates simulations of flux (at 68
frequencies from $10^{10}$ to $10^{26}$ Hz) and polarization (at radio to optical frequencies) vs.\ time. 

\begin{figure}
\includegraphics[width=75mm]{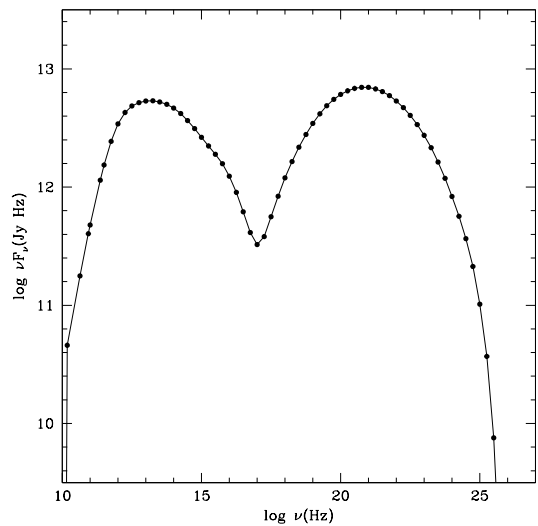}
\caption{Sample spectral energy distribution produced by the TEMZ code, with physical parameters selected
to produce emission similar to that of BL Lac during a $\gamma$-ray flare (see Fig.\ \ref{fig5}). Note the 
power-law shape in the near-IR, optical, and
near-UV range with a spectral index of 1.3, significantly steeper than both the input lower-frequency value of
$\alpha_{\rm low} = 0.55$ and the high-frequency value $\alpha_{\rm low}+0.5 = 1.05$ that one would expect solely from radiative energy losses. This results from the combination of radiative losses and the
dependence of the highest energy to which electrons are accelerated by the shock on the angle
between the magnetic field and the shock front (see text).}
\label{fig2}
\end{figure}

\begin{figure*}
\includegraphics[width=150mm]{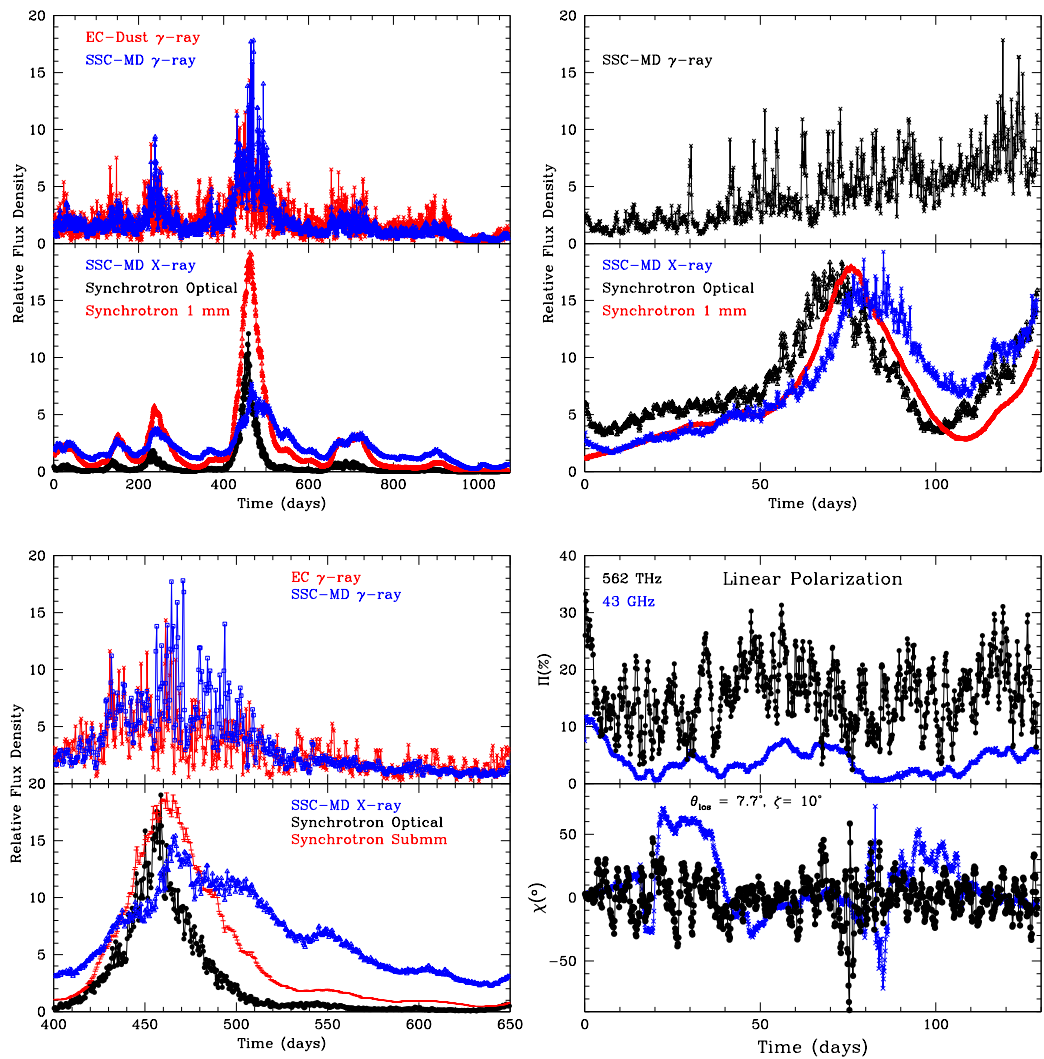}
\vspace{-5.5cm}
\caption{Sample simulated light curves and linear polarization vs. time generated by the TEMZ code. {\it Left:} Input parameters similar to those inferred observationally for 3C 454.3 \citep{wehrle12}, with
$N_{\rm cross}=60$ (see the caption of Fig.\ \ref{fig1}). Bottom panel focuses on the brightest flare seen in the top panel. {\it Right:} Input parameters similar to those inferred observationally for BL Lac \citep{mars12a}, with $N_{\rm cross}=126$. Note the changing ratios of high-energy to optical flux, sharply peaked flares, more rapid variations at higher frequencies ($\gamma$-ray vs. X-ray and optical vs. mm-wave), stronger linear polarization at higher frequencies (optical vs. mm-wave), apparent rotations in polarization position angle $\chi$, and ÒpreferredÓ polarization position angle same as jet direction ($0^\circ$ in the simulation). Some examples of ÒorphanÓ $\gamma$-ray flares are seen. All of these are observed properties of blazars.
}
\label{fig3}
\end{figure*}

The synchrotron flux density of a cell depends on (among other factors):
(1) the volume filling factor of electrons with energies high enough to radiate at the
frequency of observation,
(2) the spectral index $\alpha$, determined by the slope of the electron energy distribution,
(3) the normalization factor $N_0$ of the electron energy distribution, and
(4) the strength $B$ and angle $\psi$ (corrected for relativistic aberration) of the magnetic
field relative to the line of sight
through the factor $(B \sin \psi)^{1+\alpha}$.

Meanwhile, the inverse Compton X-ray or $\gamma$-ray flux density depends on:
(1) the volume filling factor of electrons with energies high enough to scatter
the highest-frequency seed photons to the observed photon energy,
(2) the slope of the electron energy distribution,
(3) the normalization factor $N_0$ of the electron energy distribution, and
(4) the energy density of seed photons, which is constant for photons emitted
by a dust torus and nearly constant if from the broad emission-line region, but
variable if from a relatively slow region inside the jet, such as a Mach disk.

\begin{figure}
\includegraphics[width=105mm]{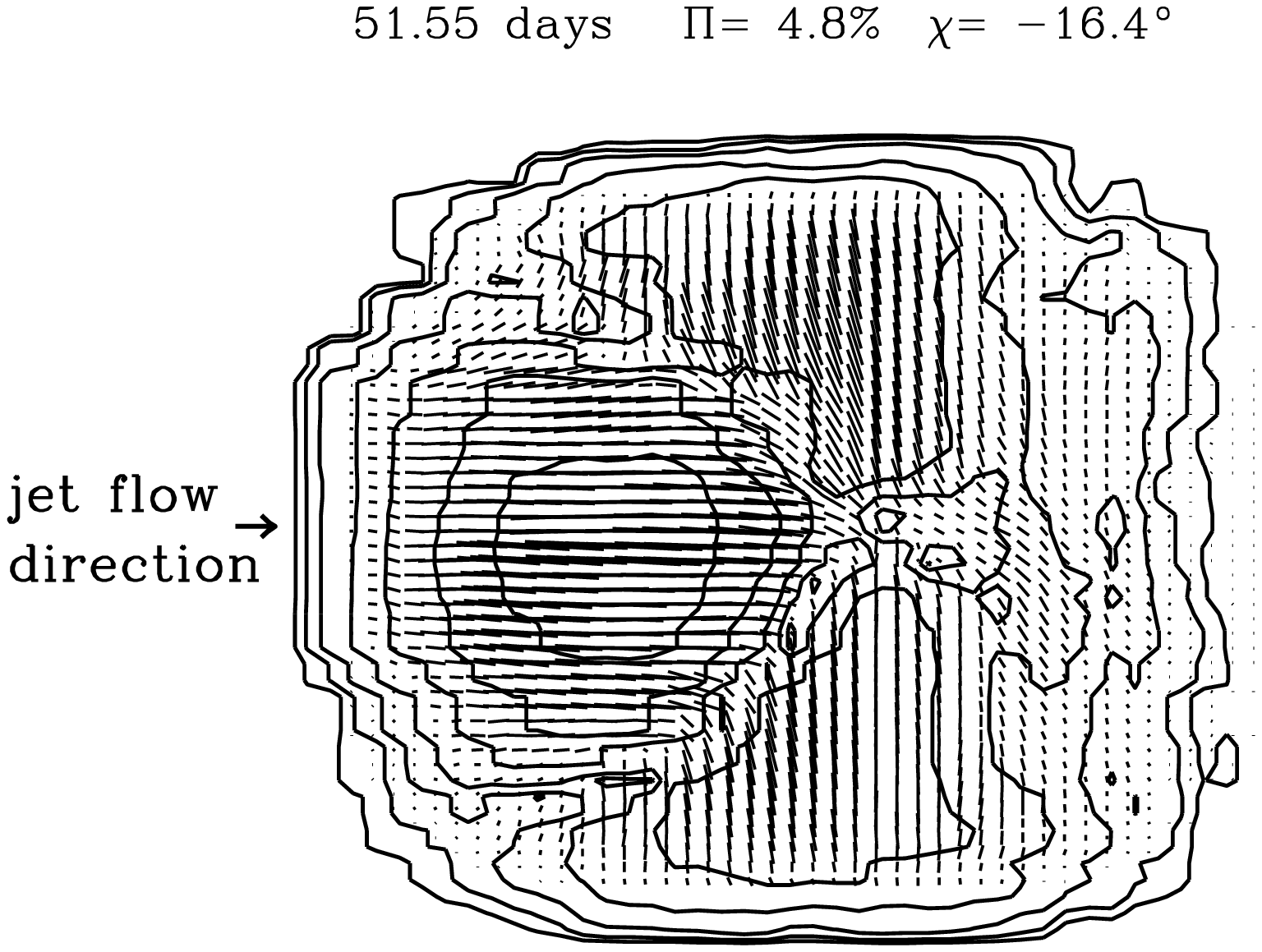}
\vspace{-3cm}
\caption{Sample polarized intensity map of the emission region between the standing shock and the
rarefaction shown in Fig.\ \ref{fig1}, at time 51.55 days in the same simulation as the one that
produced the
right-hand side of Figure \ref{fig3}. The contours are in factors of 2 starting at 1\% of the
maximum in polarization intensity. The lengths of the line segments are proportional to the
polarized intensity, and they are aligned with the polarization direction. The orientation is similar
to that of Fig.\ \ref{fig1}, with the jet base to the left. The integrated degree $\Pi$ and
position angle $\chi$ of polarization are indicated, with $\chi = 0^\circ$ parallel to the jet
axis (left-right in the figure) and positive values of $\chi$ counterclockwise to this direction.
Note that, despite the near-cancellation of the polarization from different parts of the
jet, an image of sufficiently high angular resolution would reveal strong gradients in
both the degree and position angle of polarization.}
\label{fig4}
\end{figure}

Since some of these factors are different for the two processes and for the different sources of seed photons, there will be a correlation but not a one-to-one correspondence between light curves at two widely separated frequencies despite the electron energies being roughly the same for optical and $\gamma$-ray emission. Even for relatively closely spaced frequencies (e.g., optical and near-IR), differences in volume filling factor as a function of maximum electron energy will weaken the correlations of the light curves and polarization vs. time.
At very high photon energies, ``orphan flares'' are possible if, e.g., the magnetic field of the cell that contains the highest energy electrons points directly along the line of sight. In this example, no synchrotron flare would be seen even during a major $\gamma$-ray flare.
A Mach disk at the end of the conical recollimation shock can provide a variable source of seed photons that is well beamed in the frame of the plasma passing through the recollimation shock but not in the observer's frame, since the flow speed of the plasma beyond the Mach disk is only mildly relativistic. Temporarily bright regions in a relatively slow sheath of the jet could produce a similar effect.
The code allows each cell to have a randomly oriented turbulent velocity relative to the general jet flow. This, along with the different maximum electron energy and magnetic field direction in the different cells, as well as the small volume of each cell, allows for very fast variability at optical and $\gamma$-ray frequencies \citep[see][]{np12}.

\begin{figure*}
\includegraphics[width=150mm]{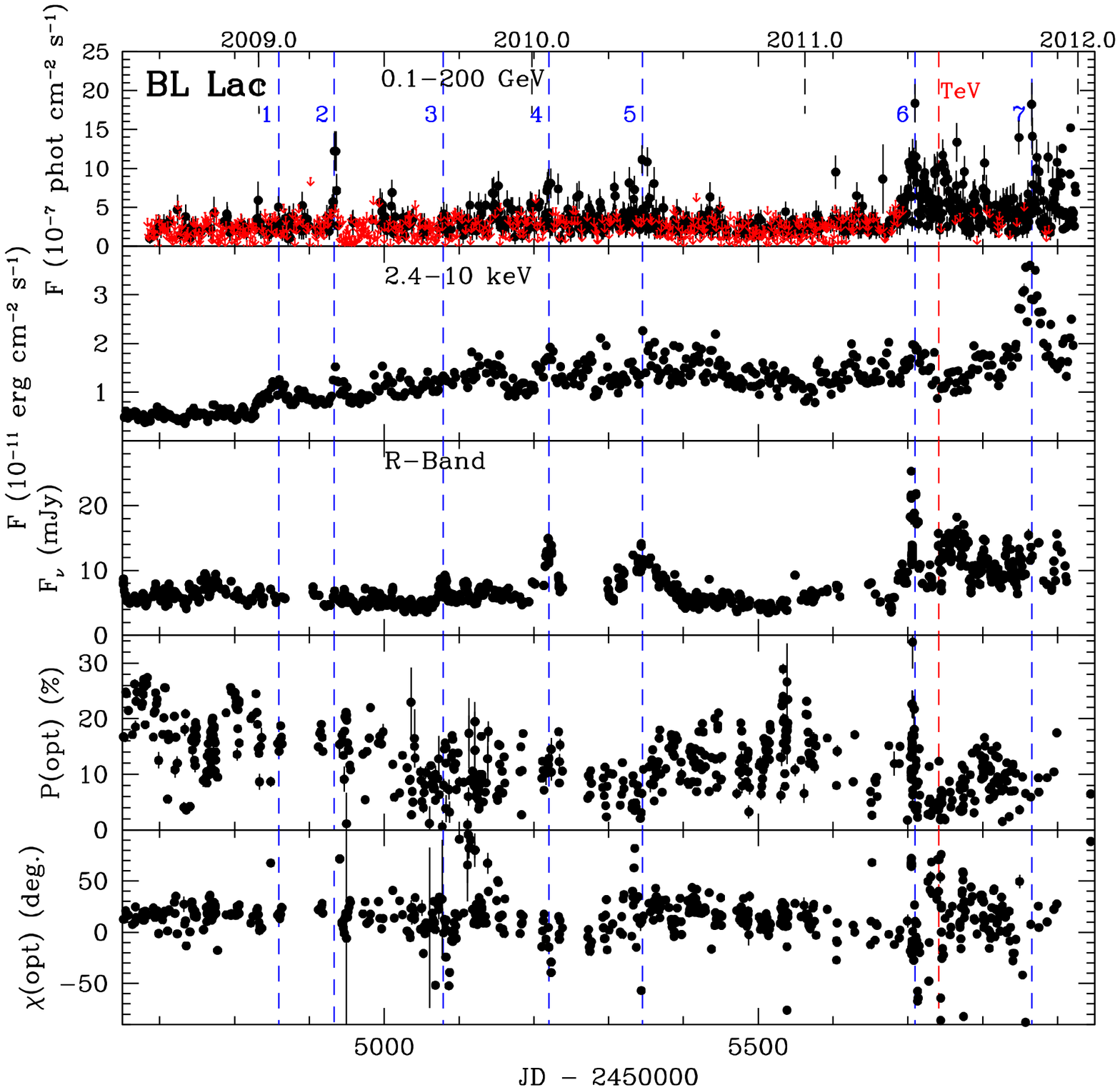}
\caption{Multi-waveband flux and degree $P$ and electric-vector
position angle $\chi$ of optical polarization vs. time for BL Lac in 2011.  
Note rapid variations in $P$ and $\chi$, indicative of a turbulent magnetic field.
Data are from Marscher et al. (in preparation).
}
\label{fig5}
\end{figure*}

If one compares the results of a simulation (Fig.\ \ref{fig3}) with actual data \citep[see Fig.\ \ref{fig5}
as well as, e.g.,][]{wehrle12}, the results look quite promising. One can see
in the plots a number of cases of apparently ``orphan'' $\gamma$-ray flares, and many flares are sharply
peaked rather than rounded, although outbursts at lower frequencies are more rounded because they
occur over larger volumes. Apparent rotations in the polarization vector occur in both clockwise
(decreasing $\chi$) and counterclockwise directions, as observed in some blazars \citep[e.g.][]{Hay12}.

Future development includes parallelization of a C++ version of the
code (which was originally written in Fortran-77) so that it can be run efficiently on a
multi-processor computer. This will allow the inclusion of SSC of time-delayed seed photons from cells
other than the Mach disk, thought to be important at least in BL Lac objects, if not quasars.
In addition, emission-line radiation from a cloud that lies next to the parsec-scale
jet \citep[see][]{LT13} will be added. This can be an additional source of time-variable
seed photons, since the excitation of the cloud will depend on the flux of UV photons emitted
by the jet.

The eventual goal is to run the code over a range of input parameters to determine the physical
conditions under which various observational features can be reproduced. Each simulation
is a unique combination of both random and systematic processes, and so will not reproduce
actual light curves. However, statistical analyses of the
timing characteristics --- correlations across frequencies, power spectra, and variations
of polarization in both time and frequency --- can be compared with the same characteristics
of the data obtained from comprehensive monitoring programs such as that led by the author's group
\citep{mar10}. In addition, ultra-high resolution polarized intensity images, such as those now
becoming available with VLBI at a wavelength of 3 mm \citep{MV12}, can verify whether the distinct
polarization structure produced by turbulent plasma crossing the standing conical shock (see
Fig.\ \ref{fig4}) is seen in nature.

\bigskip 
\begin{acknowledgments}
The author gratefully acknowledges financial support for this research from
NASA Fermi Guest Investigator grant NNX12AO79G and, during the earlier stages,
National Science Foundation grant AST-0907893. This work benefitted from International Team
collaboration 160 sponsored by the International Space Science Institute (ISSI) in
Switzerland.
\end{acknowledgments}

\bigskip 

\end{document}